\documentclass[11pt]{article}
\usepackage{amsfonts,epsfig,latexsym}

\newcommand{\R}{{\mathbb{R}}}
\newcommand{\Z}{{\mathbb{Z}}}
\newcommand{\N}{{\mathbb{N}}}

\newcommand{\beq}{\begin{equation}}
\newcommand{\eeq}{\end{equation}}
\newcommand{\bea}{\begin{eqnarray}}
\newcommand{\eea}{\end{eqnarray}}
\newcommand{\ra}{\rightarrow}

\newcommand{\cd}{\partial}

\newcommand{\osc}{C_{T,2}^+}
\newcommand{\osco}{C_{T,0}^+}
\newcommand{\om}{\Omega_{T,2}^+}
\newcommand{\omo}{\Omega_{T,0}^+}

\newcommand{\ol}{\overline}
\newcommand{\qvec}{{\bf q}}
\newcommand{\chivec}{\mbox{\boldmath{$\chi$}}}

\newtheorem{thm}{Theorem}
\newtheorem{lemma}[thm]{Lemma}

\newtheorem{defn}[thm]{Definition}

\begin{document}

\title{Breather initial profiles in chains of weakly coupled anharmonic 
oscillators}
\author{M. Haskins\thanks{E-mail: {\tt mhaskins@math.utexas.edu}} \\
Department of Mathematics, University of Texas at Austin\\
Austin, Texas 78712, U.S.A.\\ \\
J.M. Speight\thanks{E-mail: {\tt j.m.speight@leeds.ac.uk}}\\
Department of Pure Mathematics, University of Leeds\\
Leeds LS2 9JT, England}

\date{}

\maketitle
\begin{abstract}
Qualitative information about breather initial profiles in the weak coupling 
limit of a chain of identical one-dimensional anharmonic oscillators is found
by studying the linearized equations of motion at a one-site breather. In
particular, information is found about how the breather initial profile depends
on its period $T$. Numerical work shows two different kinds of breathers to exist
and to occur in alternating $T$-bands. Genericity of certain aspects of the
observed behaviour is proved. 

\end{abstract}

\section{Introduction}
\label{int}

Chains of coupled anharmonic
oscillators have many applications in condensed matter and
biophysics as simple one-dimensional models of crystals or biomolecules. 
Of particular interest are the so-called ``breather''
solutions supported by such chains, that is, oscillatory solutions which
are periodic in time and exponentially localized in space \cite{review}. 
The simplest 
possible class of models, where all the oscillators are identical,
with one degree of freedom, and 
nearest neighbours are coupled by identical Hooke's law springs has been
widely studied in this context. The equation of motion for the position
$q_n(t)$ of the $n$-th oscillator ($n\in\Z$) is
\beq
\label{1}
\ddot{q}_n-\alpha(q_{n+1}-2q_{n}+q_{n-1})+V'(q_n)=0
\eeq
where $\alpha$ is the spring constant and
 $V$ is the anharmonic substrate potential, which we choose to normalize
so that $V'(0)=0$ and $V''(0)=1$.
MacKay and Aubry have proved the existence of breathers in the weak coupling
(small $\alpha$) regime of this system \cite{macaub}. They noted that in the
limit $\alpha\ra 0$, system (\ref{1}) supports a one-site breather, call it
$\qvec_0=(q_n)_{n\in\Z}$, where one site ($n=0$ say) oscillates with period
$T$ while all the others remain stationary at $0$. Using an implicit function
theorem argument they proved existence of a constant period continuation $\qvec_\alpha$ of 
this breather away from $\alpha=0$ provided $\alpha$ remains sufficiently
small and $T\notin 2\pi\Z^+$.

Having established the existence of these breathers, the question remains:
what do they look like? This question may be addressed by means of numerical
analysis in several ways. One technique is to seek period $T$ solutions of
a spatially truncated (i.e. $|n|\leq N<\infty$) version of (\ref{1}) by 
searching for fixed points of the period $T$ Poincar\'e return map using, for
example, a Newton-Raphson algorithm \cite{aubmar}. This method potentially
allows the construction of breathers far from $\alpha=0$, and can be used to
determine the domain of existence of continued one-site breathers in the 
$(\alpha,T)$ parameter space (as applied by the authors to a non-standard
discrete sine-Gordon system in \cite{hasspe}). This technique has its 
drawbacks, however. It is rather computationally expensive and there are
subtleties concerning convergence of the Newton-Raphson method close to
$\alpha=0$ \cite{aubmar,hasspe}. 
On the other hand, if the weak coupling regime is of primary
interest, then much useful information may be obtained from
\beq
\qvec'_0:=\left.\frac{\cd\qvec_\alpha}{\cd\alpha}\right|_{\alpha=0},
\eeq
the tangent vector at $\alpha=0$ to the continuation curve $\qvec_\alpha$.
This vector may be calculated by solving a certain initial value problem for
a set of 4 coupled second order ODEs, a numerically trivial task. The small
$\alpha$ behaviour of breathers can then be determined by approximating the
curve $\qvec_\alpha$ by its tangent line at $\alpha=0$, that is
\beq
\label{2}
\qvec_\alpha=\qvec_0+\alpha\qvec'_0+o(\alpha).
\eeq
Since this method is so computationally cheap, it is easy to make a 
systematic study of the period dependence of breather initial profiles for a 
variety of substrate potentials. This paper presents such a study.

The paper is organized as follows. In section \ref{dir}
 we give a precise statement
of the MacKay-Aubry existence theorem and reduce the calculation of
$\qvec'_0$ to numerically tractable form. In section \ref{num} we present
numerically generated graphs of the components of $\qvec'_0$ against period
$T$ for various choices of potential, and extract from them information about
breather initial profiles in the small $\alpha$ regime. The results show
a generic behaviour of alternating $T$-bands of two qualitatively different
types of breather, which we call ``in-phase breathers'' and 
``anti-phase breathers''. In section \ref{gen} we prove that this alternating
behaviour is generic in a precise sense, and examine the small $T$ limit
analytically. Section \ref{con} contains some concluding remarks.

\section{The direction of continuation of one-site breathers}
\label{dir}

In the following we will assume that $V:\R\ra\R$ is twice continuously
differentiable and 
has a normalized stable equilibrium point at $0$ ($V'(0)=0$, $V''(0)=1$).
Consider the equation of motion for a particle moving in such a potential,
\beq
\label{3}
\ddot{x}+V'(x)=0
\eeq
with $\dot{x}(0)=0$. Provided $|x(0)|$ is small enough, $x(t)$ must be a
periodic oscillation. All the potentials we consider will be 
anharmonic
with classical spectrum $(2\pi,\infty)$. Anharmonic means that the period of
oscillation varies nondegenerately with $x(0)>0$. The classical 
spectrum of $V$ is the set of periods of the oscillations supported by
$V$. 

Let $x_T(t)$ denote the solution of 
(\ref{3}) with period $T>2\pi$ and $x(0)>0$ which has even time-reversal
symmetry, $x_T(-t)\equiv x_T(t)$. From this we may construct a 1-site 
breather solution of system (\ref{1}) with $\alpha=0$, call it $\qvec_0$:
\beq
\label{4}
q_{n,0}(t)=\left\{
\begin{array}{cc}
x_T(t) & n=0 \\
0 & n\neq 0.
\end{array}\right. 
\eeq
The MacKay-Aubry theorem establishes the existence of a continuation 
$\qvec_\alpha$ of this solution away from $\alpha=0$ in a suitable function
space, defined as follows.

\begin{defn} For any $n\in\N$, 
let $C^+_{T,n}$ denote the space of $n$ times 
continuously differentiable mappings $\R\ra\R$ which are $T$-periodic
and have even time reversal symmetry. Note that $C_{T,n}^+$ is a
Banach space when equipped with the uniform $C^n$ norm:
$$
|q|_n:=\sup_{t\in\R}\{|q(t)|,|\dot{q}(t)|,\ldots,|q^{(n)}(t)|\}.
$$
\end{defn}

\begin{defn} For any $n\in\N$,
$$
\Omega_{T,n}^+:=\{\qvec:\Z\ra C^+_{T,n}\,\, \mbox{\rm such that}\,\,
 ||\qvec||_n<
\infty\}
$$
where
$$
||\qvec||_n:=\sup_{m\in\Z}|q_m|_n.
$$
Note that $(\Omega_{T,n}^+,||\cdot||_n)$ is also a Banach space.
\end{defn}

The required function space is $\om$.
By construction, $\qvec_0\in\om$ and is exponentially spatially localized.

\begin{thm}(MacKay-Aubry) 
\label{th:macaub}
If $T\notin 2\pi\Z$ there exists $\epsilon>0$
such that for all $\alpha\in[0,\epsilon)$ there is a unique continuous
family $\qvec_\alpha\in\om$ of solutions of system (\ref{1}) at
coupling $\alpha$ with 
$\qvec_0$ as defined in (\ref{4}). These solutions are 
exponentially localized in space and the map $[0,\epsilon)\ra\om$ given by
$\alpha\mapsto\qvec_\alpha$ is $C^1$. 
\end{thm}

The idea of the proof is to define a $C^1$ mapping $F:\om\oplus\R\ra\omo$,
\beq
F(\qvec,\alpha)_m=\ddot{q}_m-\alpha(q_{m+1}-2q_m+q_{m-1})+V'(q_m),
\eeq
so that $F(\qvec,\alpha)=0$ if and only if $\qvec$ is an even $T$-periodic
 solution of system (\ref{1}). In particular, $F(\qvec_0,0)=0$ by 
construction. Using $T\notin 2\pi\Z^+$ and anharmonicity of $V$, one can show
that the partial derivative of $F$ with respect to $\qvec$ at $(\qvec_0,0)$,
$DF_{q_0}:\om\ra\omo$, is invertible ($DF_{q_0}$ is injective with 
$(DF_{q_0})^{-1}$ bounded).
Hence the implicit function theorem \cite{chobru1} applies and local 
existence and uniqueness of the $C^1$ family $\qvec_\alpha$ satisfying
\beq
\label{4.5}
F(\qvec_\alpha,\alpha)=0
\eeq
 are assured. Persistence of exponential 
localization is proved as a separate step.

The object of interest in this paper is $\qvec'_0\in\om$, the tangent vector 
to the curve $\qvec_\alpha$ at $\alpha=0$. This may be constructed by
implicit differentiation of (\ref{4.5}) with respect to $\alpha$ at $\alpha
=0$:
\bea
DF_{q_0}\, \qvec_0'+\left.\frac{\cd F}{\cd\alpha}\right|_{(\qvec_0,0)}&=&0
\nonumber \\
\label{5}
\Rightarrow\qquad
\qvec_0'&=&-(DF_{q_0})^{-1}\, 
\left.\frac{\cd F}{\cd\alpha}\right|_{(\qvec_0,0)}.
\eea
Straightforward calculation shows that
\beq
\label{6}
[DF_{q_0}\, \chivec]_m=\left\{
\begin{array}{cc}
\ddot{\chi}_m+\chi_m & m\neq 0 \\
\ddot{\chi}_m+V''(x_T)\chi_m & m=0
\end{array}\right.
\eeq
for any $\chivec\in\om$, and
\beq
\label{6.5}
\left(\left.\frac{\cd F}{\cd\alpha}\right|_{(\qvec_0,0)}\right)_m=
\left\{\begin{array}{cc}
2x_T & m=0 \\
-x_T & |m|=1 \\
0 & |m|>1.
\end{array}\right.
\eeq
So evaluating the right hand side of (\ref{5}) to find $\qvec_0'=\chivec
\in\om$ is equivalent to solving the following infinite decoupled set of
ODEs for $\{\chi_m\in\osc:m\in\Z\}$:
\bea
\label{7}
\ddot{\chi}_0+V''(x_T)\chi_0&=&-2x_T \\
\label{8}
\ddot{\chi}_m+\chi_m&=&x_T\qquad\qquad |m|=1 \\
\label{9}
\ddot{\chi}_m+\chi_m&=&0\qquad\qquad\,\,\,\,\, |m|>1.
\eea
Recalling that $T\notin 2\pi\Z^+$ (so $\cos\notin\osc$) one sees from 
(\ref{9}) that $\chi_m\equiv 0$ for $|m|>1$, so one need only compute
$\chi_0$ and $\chi_1$ (clearly $\chi_{-1}\equiv\chi_1$).

In each case we must find the even $T$-periodic solution of an inhomogeneous
2nd order linear ODE.\, 
For $m=0,1$ let $y_m(t)$ be the solution
of the corresponding homogeneous equation with initial data $y_m(0)=1$, $\dot{y}_m(0)=0$,
so
\bea
\label{10}
\ddot{y}_0+V''(x_T)y_0&=&0 \\
\label{11}
\ddot{y}_1+y_1&=&0. 
\eea
(Of course $y_1=\cos$.) Similarly, let $z_m(t)$ denote the particular 
integral of the inhomogeneous equation with initial data $z_m(0)=\dot{z}_m(0)=0$, so
\bea 
\label{12}
\ddot{z}_0+V''(x_T)z_0&=&-2x_T \\
\label{13}
\ddot{z}_1+z_1&=&x_T.
\eea
Since $\chi_m(t)$ is even, it follows that $\chi_m(t)=\chi_m(0)y_m(t)+
z_m(t)$, and $\chi_m(0)$ may be determined by applying either of the
$T$-periodicity constraints
\bea
\label{14}
\chi_m(T)=\chi_m(0) &\Rightarrow& \chi_m(0)=\frac{z_m(T)}{1-y_m(T)} \\
\label{15}
\mbox{or}\qquad
\dot{\chi}_m(T)=\dot{\chi}_m(0) &\Rightarrow& \chi_m(0)=
-\frac{\dot{z}_m(T)}{\dot{y}_m(T)}.
\eea
In either case, one sees that the tangent vector $\qvec_0'$, and in 
particular its initial value $\qvec_0'(0)=(\ldots,0,0,\chi_1(0),\chi_0(0),
\chi_1(0),0,0,\ldots)$ can be constructed once the 4 initial value problems
(\ref{10}--\ref{13}) are solved on $[0,T]$.

Having computed the initial value of the tangent vector, we may approximate
the breather initial profile for small $\alpha$ using (\ref{2}):
\beq
\label{16}
q_{m,\alpha}(0)=\left\{
\begin{array}{ccccccl}
x_T(0) & + & \alpha\chi_0(0) & + & o(\alpha) & & m=0 \\
       &   & \alpha\chi_1(0) & + & o(\alpha) & & |m|=1 \\
       &   &                 &   & o(\alpha) & & |m|>1.
\end{array}
\right.
\eeq
So to first order in $\alpha$, the continuation leaves all but the central
($m=0$) and off-central ($m=\pm 1$) sites at the equilibrium position. The
qualitative shape of the breather initial profile depends crucially on
$\chi_1(0)$ (but not $\chi_0(0)$). If $\chi_1(0)>0$, the continuation 
displaces the off-central sites from equilibrium in the same direction as
the central site. The result is a hump shaped breather in which the central
and off-central sites oscillate, roughly speaking, in phase (that is they
attain their maxima and minima simultaneously). We shall call such breathers
``in-phase breathers (IPBs).'' If $\chi_1(0)<0$, on the other hand, the
off-central sites are displaced in the opposite direction from the central 
site, resulting in a sombrero shaped initial profile. In this case, the
central and off-central sites oscillate, roughly speaking, in anti-phase
(the central site attains its maximum when the off-central sites attain their
minima, and vice-versa). We shall call such breathers ``anti-phase breathers
(APBs).'' Sievers and Takeno \cite{sietak} have argued (without giving a
rigorous mathematical proof)
that breathers of the 
latter type are supported by certain
oscillator chains with no substrate potential and anharmonic nearest 
neighbour
coupling, but their existence in systems of type (\ref{1}) does not seem to
have attracted attention previously in the literature.
This is surprising in light of the numerical results
described in section \ref{num}. We shall find that for a generic substrate
potential $V$, system (\ref{1}) supports both IPBs and APBs, the type
varying in bands as $T$ increases through $(2\pi,\infty)$.

\section{Numerical results}
\label{num}

Calculation of the constants $\chi_0(0)$ and $\chi_1(0)$ may be performed
numerically by solving the initial value problems (\ref{10}), (\ref{12}) and
(\ref{13}) using some approximate ODE solver (recall the exact solution of
(\ref{11}) is known). The results presented in this section were generated
using a 4th order Runge-Kutta method with fixed time step $\delta t=0.01$.
Each ODE requires the periodic oscillation $x_T(t)$ as input. For most
potentials this function must itself be generated by numerical solution
of the oscillator equation (\ref{3}), which might just as well be solved in
parallel with (\ref{10},\ref{12},\ref{13}), making 4 coupled 2nd order
ODEs in all.
Since the correct initial displacement $x_T(0)$ for a given period $T$ is
not known in advance, the initial value problem is parametrized by initial
displacement, $T$ being determined from the approximate solution of (\ref{3})
by counting sign changes of $x_T(t)$ and linear interpolation.

The results obtained depend crucially on whether the substrate potential
has reflexion symmetry about the equilibrium position.

\subsection{Asymmetric potentials}
\label{asy}

The following asymmetric potentials were investigated:
\bea
\label{17}
\mbox{Morse:} &V_M(x)& =\frac{1}{2}(1-e^{-x})^2 \\
\label{18}
\mbox{Lennard-Jones:} &V_{LJ}(x)&=\frac{1}{72}\left(\frac{1}{x^{12}}-
\frac{2}{x^6}\right) \\
\label{19}
\mbox{Cubic:} &V_C(x)&=\frac{1}{2}x^2-\frac{1}{3}x^3.
\eea
Note that the equilibrium position for $V_{LJ}(x)$ is $x=1$ rather than 
$x=0$, as we have been assuming so far. Hence, even though $V_{LJ}$ is even,
it is still asymmetric, since it is not reflexion symmetric about $x=1$.
Graphs of $\chi_0(0)$ and $\chi_1(0)$ against $T$ for all these potentials
are presented in  figures 1 and 2 respectively.

\vbox{
\centerline{\epsfysize=3truein
\epsfbox[63   200   549   589]{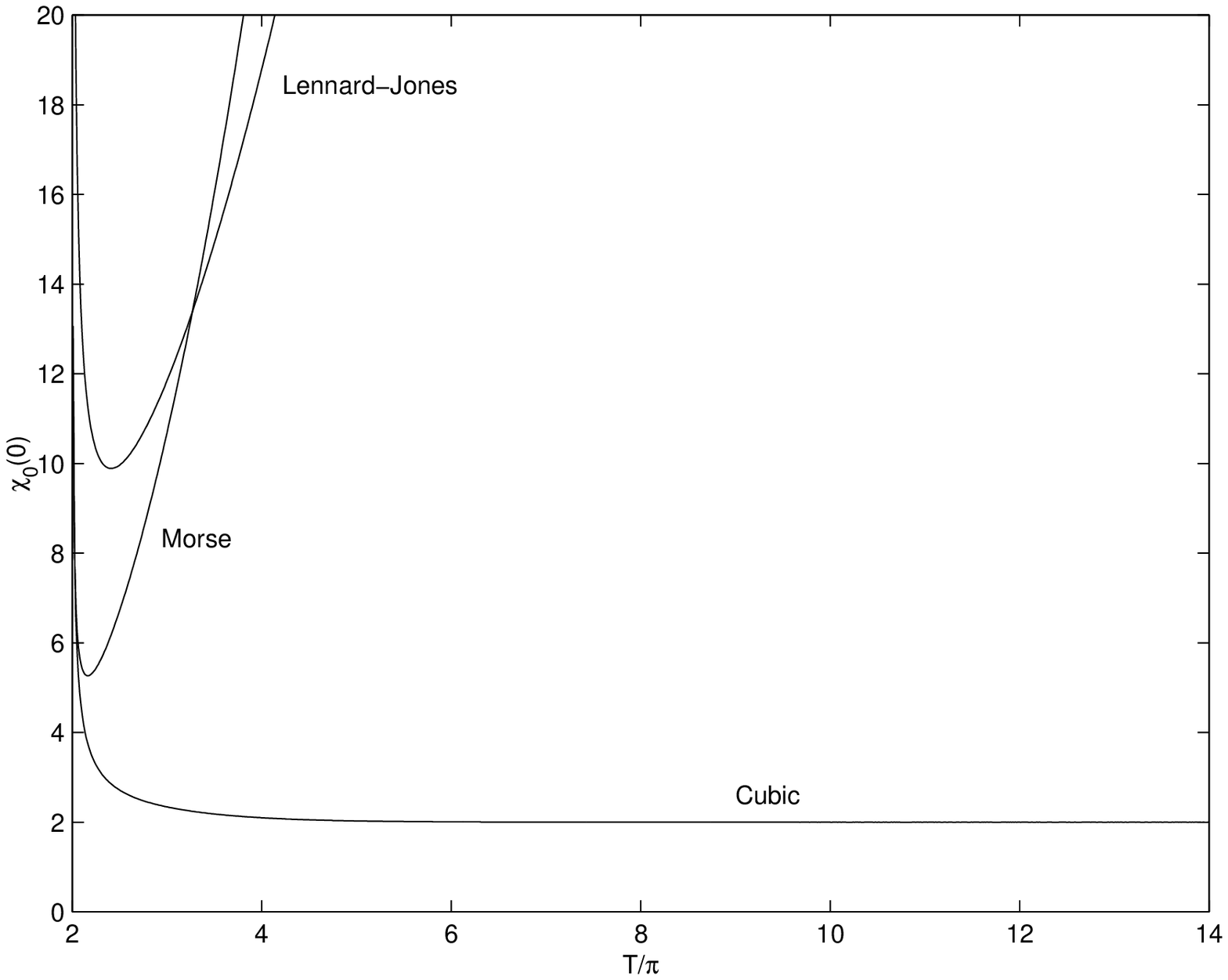}}
\noindent
{\it Figure 1: 
Graphs of $\chi_0(0)$ against $T$ for various asymmetric substrate 
potentials.}
}
\vspace{0.5cm}

Figure 1 illustrates a clear difference in the
large $T$ behaviour of $\chi_0(0)$ between $V_C$ and the other two potentials
$V_M$ and $V_{LJ}$. Namely $\chi_0(0)$ remains bounded as $T\ra\infty$ for
$V_C$, but grows unbounded for $V_M$ and $V_{LJ}$. This phenomenon appears
to be linked to the large $T$ behaviour of the $T$ periodic oscillations
$x_T(t)$. Due to the unstable equilibrium point of $V_C$ at $x=1$, $x_T(t)$
tends to a bounded homoclinic orbit as $T\ra\infty$ for this potential, and
$\chi_0(0)$ is correspondingly bounded for large $T$. For
$V_M$ and $V_{LJ}$ no such bounded
homoclinic orbit exists: $x_T(0)$ is an unbounded
function of $T$ and $\chi_0(0)$ is correspondingly unbounded as $T\ra\infty$.
In all cases, we note that $\chi_0(0)>0$, so the continuation to $\alpha>0$
displaces the central site further from equilibrium.

\vbox{
\centerline{\epsfysize=3truein
\epsfbox[63   200   549   589]{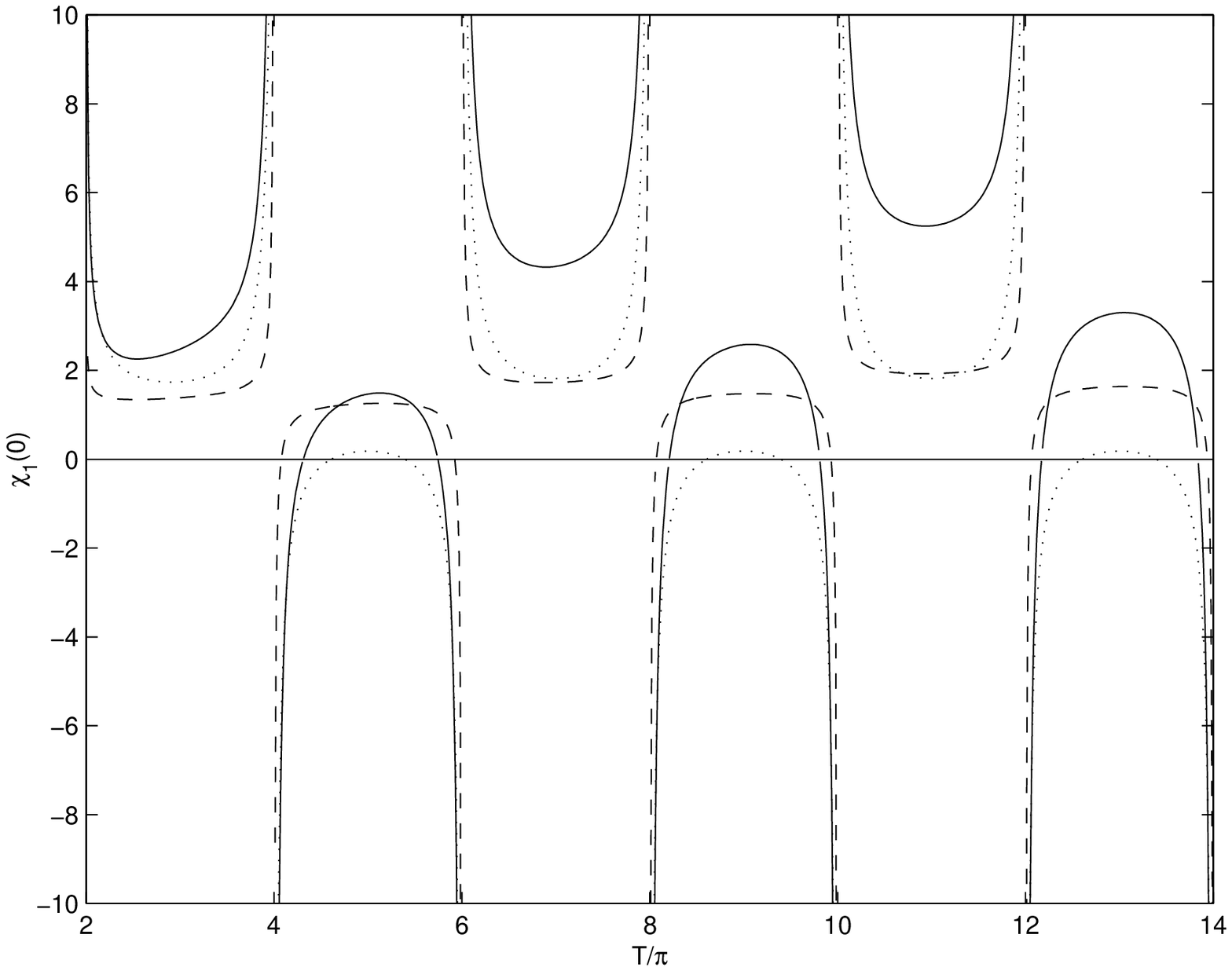}}
\noindent
{\it Figure 2: 
Graphs of $\chi_1(0)$ against $T$ for various asymmetric substrate 
potentials (solid: Morse; dashed: Lennard-Jones; dotted: cubic).}
}
\vspace{0.5cm}

From figure 2 we see that the sign of $\chi_1(0)$ changes as 
$T$ increases, leading to the prediction of
$T$-bands of IPBs and APBs, as explained in section \ref{dir}. The sign 
changes are clearly associated with the vertical asymptotes of the
graphs at each $T\in 2\pi\Z^+$ (note that these do not conflict with the
contents of section \ref{dir} since the existence proof for $\qvec_\alpha$,
and hence for $\qvec'_0$, breaks down when $T\in 2\pi\Z^+$). The presence
of such asymptotes may be understood by using Green's function techniques
to write down $z_1(t)$, the solution of (\ref{13}):
\beq
\label{20}
z_1(t)=\int_0^t\sin(t-s)x_T(s)\, ds.
\eeq
Combining (\ref{20}) with (\ref{15}), and recalling that
$y_1=\cos$, yields a formula for $\chi_1(0)$,
\beq
\label{22}
\chi_1(0)=\frac{1}{\sin T}{\int_0^T\cos(T-t)x_T(t)\, dt}.
\eeq
Equation (\ref{22}) shows that there is a vertical asymptote at $T=2n\pi$,
with a sign change in $\chi_1(0)$, unless
\beq
\label{23}
\int_0^{2n\pi}\cos t\, x_{2n\pi}(t)\, dt=0,
\eeq
that is, unless $x_{2n\pi}(t)$ has vanishing $n$-th Fourier coefficient.
It might appear from (\ref{22}) that there should also be asymptotes at
$T=(2n+1)\pi$. Of course, this cannot be true since standard results
on continuity of solutions of ODEs with respect to initial data imply that
$\chi_1(0)$ is continuous 
for $T\notin 2\pi\Z^+$. In fact, a simple argument using
periodicity and evenness of $x_T$ demonstrates that
\beq
\label{24}
\int_0^{(2n+1)\pi}\cos t\, x_{(2n+1)\pi}(t)\, dt\equiv 0
\eeq
for all $n\in\Z^+$ and $V$. By contrast, we will prove in section \ref{gen}
that (\ref{23}) almost never holds (in a sense which will be made precise),
so that the resonant periods $T=2n\pi$ generically separate IPB bands from
APB bands.

\subsection{Symmetric potentials}
\label{sym}

The following symmetric potentials were investigated:
\bea
\mbox{Frenkel-Kontorova:} & V_{FK}(x) & =1-\cos x \\
\mbox{Quartic:}           & V_Q(x)    & =\frac{1}{2}x^2-\frac{1}{4}x^4 \\
\mbox{Gaussian:}          & V_G(x)    & =1-e^{-x^2/2}
\eea
The graphs in figure 3 show $\chi_0(0)$ against $T$ and closely 
resemble those of the previous section: $\chi_0(0)$ is bounded as $T\ra\infty$ if $x_T$ tends
to a bounded homoclinic orbit ($V_{FK}$ and $V_Q$), and unbounded if
$x_T(0)\ra\infty$ as $T\ra\infty$ ($V_G$).

\vbox{
\centerline{\epsfysize=3truein
\epsfbox[63   200   549   589]{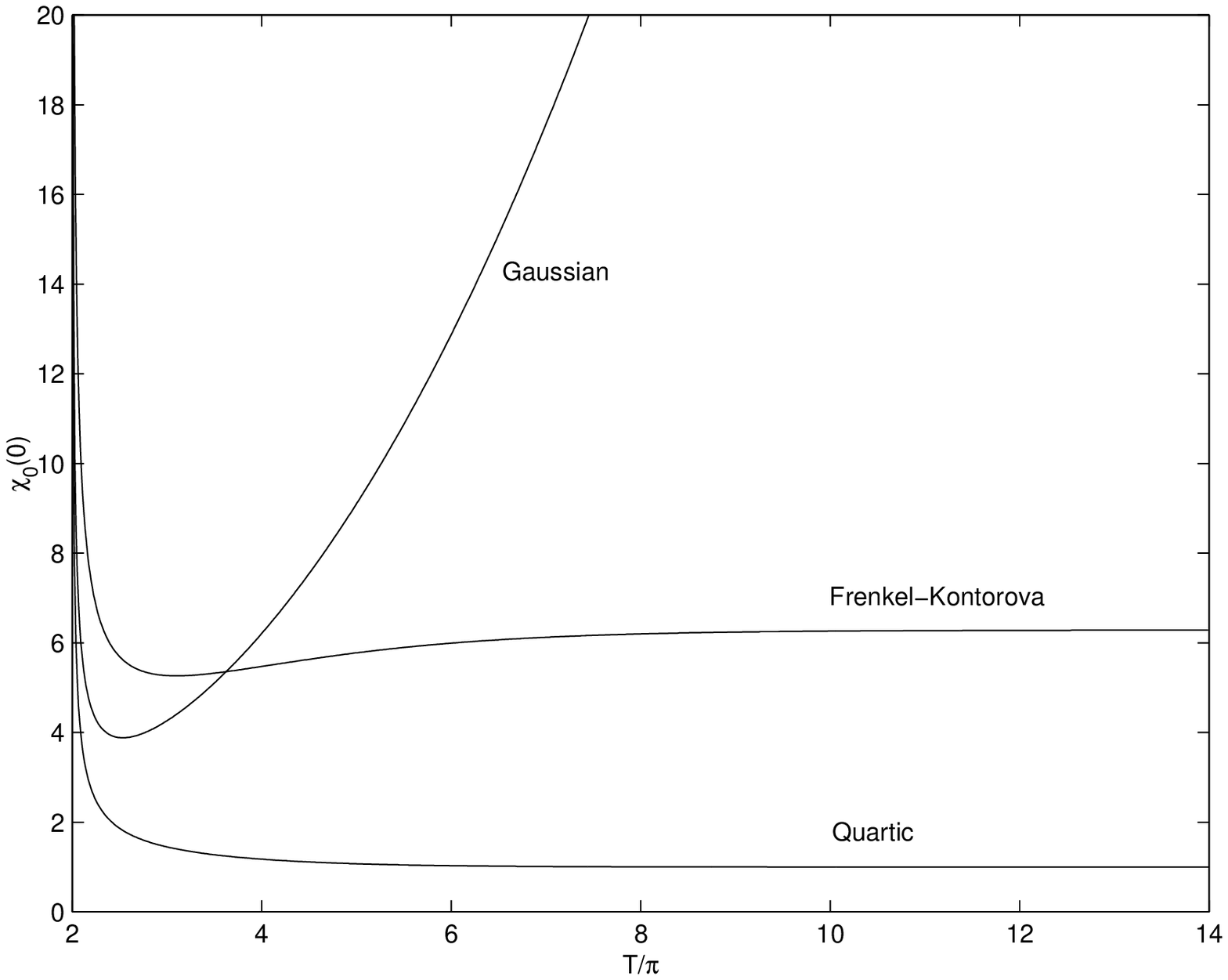}}
\noindent
{\it Figure 3: 
Graphs of $\chi_0(0)$ against $T$ for various symmetric substrate 
potentials.}
}
\vspace{0.5cm}

On the other hand, the graphs of $\chi_1(0)$ against $T$ (see figure 4) 
are strikingly different from the asymmetric case. In each case, vertical
asymptotes are present at $T=2n\pi$ only if $n $ is an {\em even} positive
integer. If $n$ is odd, no asymptote is present. To explain this, note that
evenness of $V$ implies that $x_T(t)$ is $T/2$ antiperiodic, that is
$x_T(t-T/2)\equiv -x_T(t)$, which in turn guarantees that equation (\ref{23})
holds whenever $n$ is odd. The limit $\lim_{T\ra 2n\pi}\chi_1(0)$ ($n$ odd)
exists and may, in principle, be computed using L'Hospital's rule with
\beq
\frac{d\,\,}{dT}z_1(T)
=x_T(T)+\int_0^T[-\sin(T-t)x_T(t)+\cos(T-t)f'(T)y_0(t)]\, dt,
\eeq
where $y_0$ is the solution of (\ref{10}) as before and $f(T):=x_T(0)$.
 
\vbox{
\centerline{\epsfysize=3truein
\epsfbox[63   200   549   589]{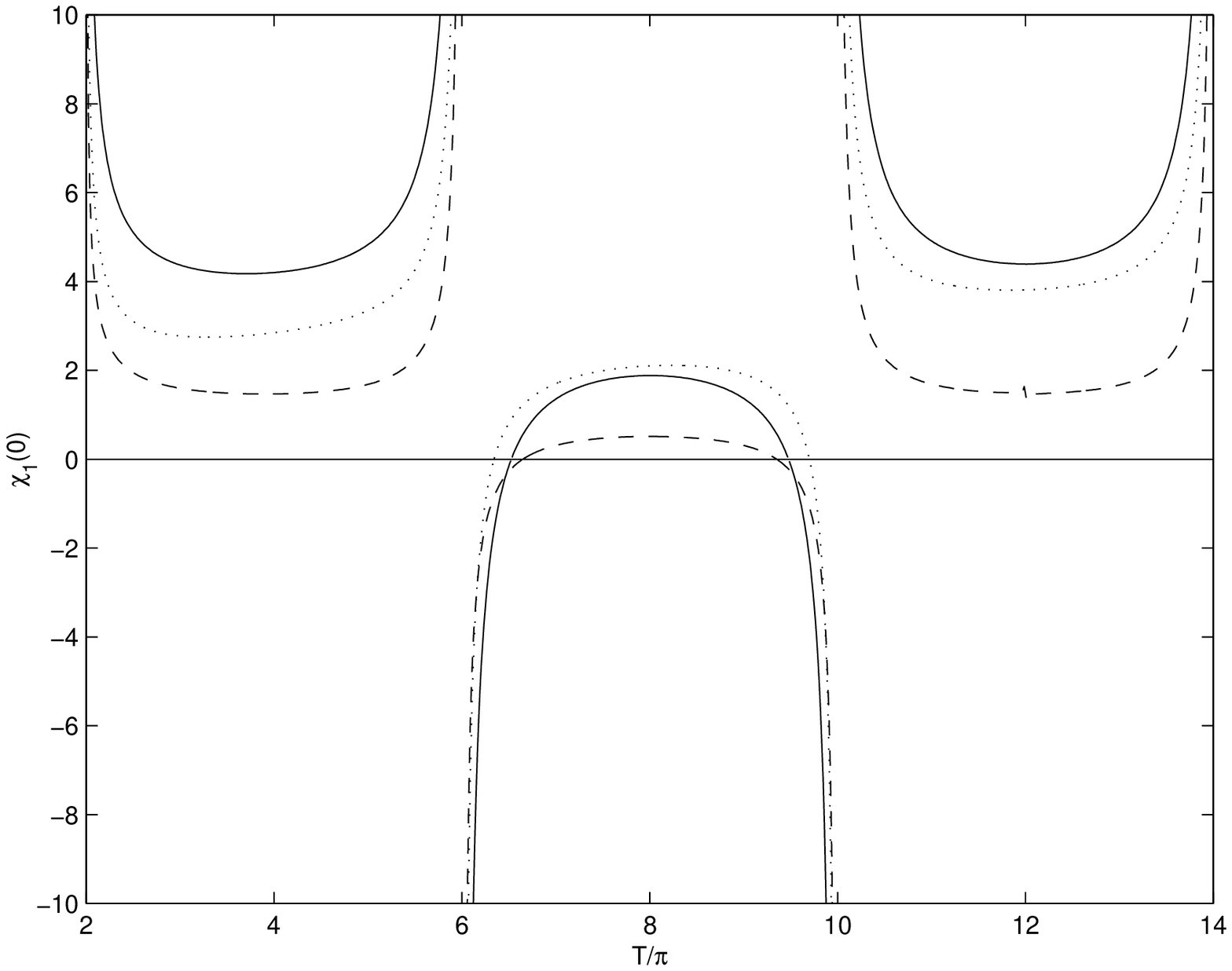}}
\noindent
{\it Figure 4: 
Graphs of $\chi_1(0)$ against $T$ for various symmetric substrate 
potentials (solid: Frenkel-Kontorova; dashed: quartic; dotted: Gaussian).}
}
\vspace{0.5cm}

\subsection{Almost symmetric potentials}
\label{alm}

For $m=1$ equation (\ref{14}) becomes
\beq
\label{14one}
\chi_1(0)= \frac{z_1(T)}{1-\cos{T}}. 
\eeq
From this equation it is clear that the behaviour of $\chi_1(0)$ as a 
function of $T$ is largely determined by the zeroes of $z_1(T)$. In
particular, the sign of $z_1(T)$ determines the sign of $\chi_1(0)$. 
Figure 5 shows typical zero distributions for both 
asymmetric and symmetric potentials. It is interesting to note that no 
infinitesimal perturbation of the symmetric type distribution yields the
asymmetric type distribution. Hence the dependence of $\chi_1(0)$ on $T$
for an almost symmetric
potential must differ qualitatively from the typical asymmetric
behaviour observed in section \ref{asy}.
\vspace{0.5cm}

\vbox{
\centerline{\epsfysize=2.0truein
\epsfbox[95 345 499 497]{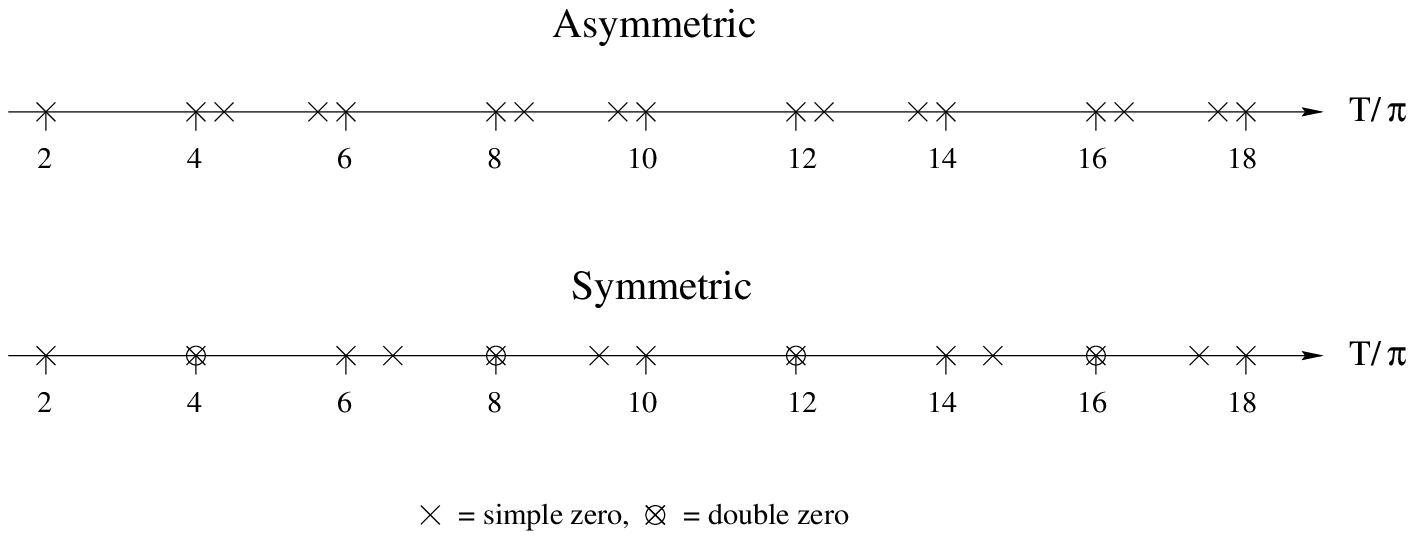}}
\noindent
{\it Figure 5: 
Typical zero distributions of $z_1(T)$ for asymmetric and symmetric
substrates.}
}

This observation motivates us to consider $z_1(T)$ for the following 
1-parameter family of potentials:
\beq
\label{29}
V_\epsilon(x)=\frac{1}{2}x^2-\frac{1}{4}x^4-\frac{\epsilon}{5}x^5.
\eeq
The zeroes of $z_1(T)$ in a region of the $(T,\log\epsilon)$ plane are 
plotted in figure 6. In the symmetric limit ($\log\epsilon\ra
-\infty$), the typical symmetric type distribution is recovered, as expected.
For a fixed but very small $\epsilon>0$ (e.g.\ $\log\epsilon=-30$), 
the small $T$ oscillations $x_T(t)$
are pointwise essentially unchanged from the $\epsilon=0$ case, and the
distribution for small $T$ is numerically indistinguishable from the
symmetric case. Only for large $T$ does the oscillation $x_T(t)$ detect the
asymmetry of $V_\epsilon$, and a transition from symmetric to asymmetric
behaviour occurs in the distribution. 
As $\epsilon$ is increased, the period
at which this transition occurs becomes smaller and smaller, so that the
transition ``propagates'' leftwards as seen in figure 6. For
sufficiently large $\epsilon$ (e.g.\ $\log\epsilon>-3$) 
the transition is complete and the typical
asymmetric distribution is recovered. These observations appear to be 
essentially independent of the asymmetric perturbation considered.
\vspace{0.5cm}

\vbox{
\centerline{\epsfysize=3truein
\epsfbox[63   200   549   589]{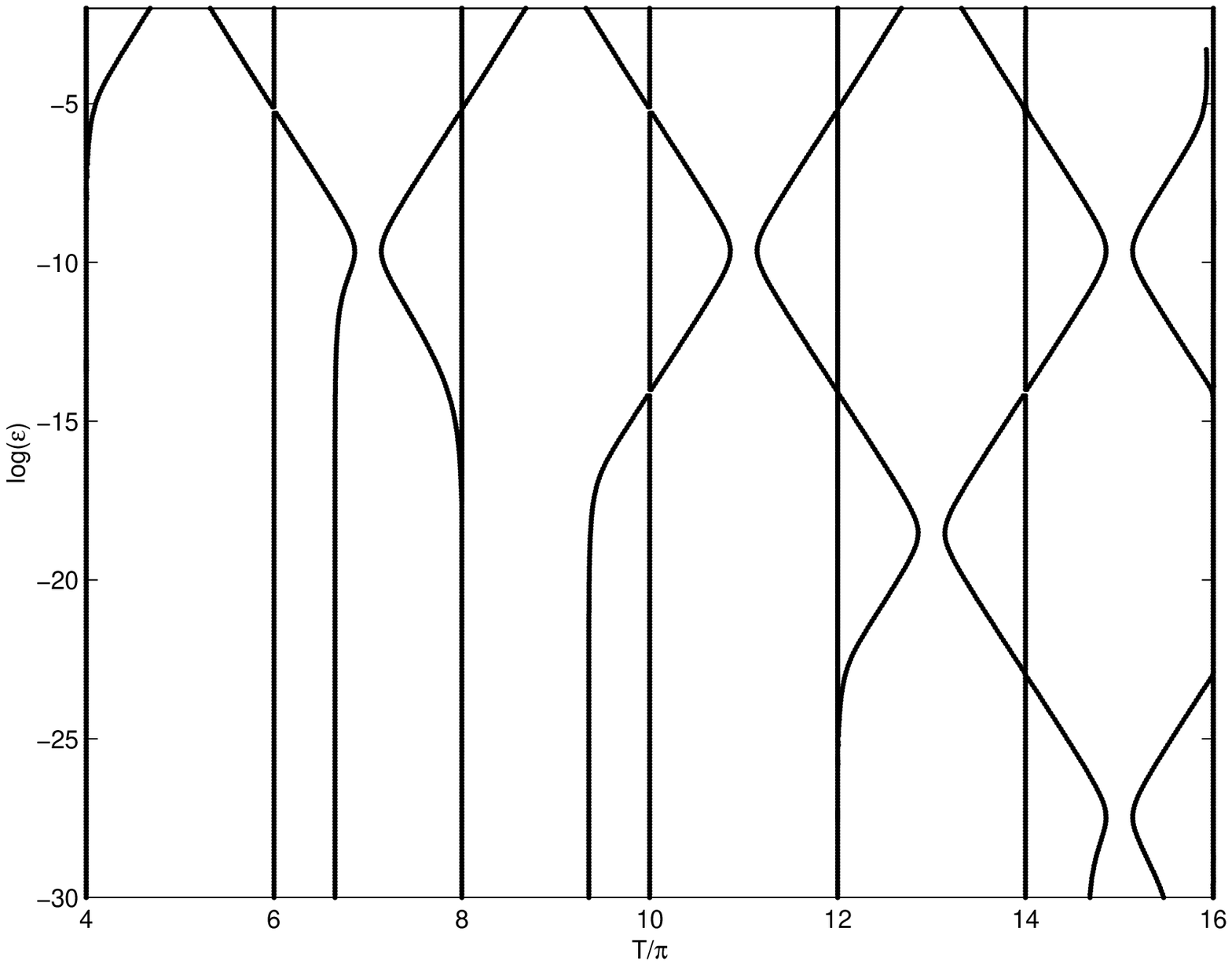}}
\noindent
{\it Figure 6: 
Zero distribution of $z_1(T)$ for the one parameter family of
almost symmetric substrates $V_\epsilon(x)=\frac{1}{2}x^2-\frac{1}{4}x^4
-\frac{\epsilon}{5}x^5$.}
}

\section{Genericity of numerical results}
\label{gen}

The aim of this section is to explain some of the generic features observed
in the numerical results of section \ref{num}. In particular we will
consider the $T\ra 2\pi^+$ behaviour of $\chi_1(0)$, and the generic presence
of asymptotes at $T\in 2\pi\Z^+$.

\subsection{The small period limit}
\label{sma}

The numerical data suggest that $\chi_1(0)\ra+\infty$ as $T\ra 2\pi^+$ for
every potential $V$. We may confirm this prediction, at least in the case 
where $V$ is analytic (as is the case in all our examples), by means of a 
Linstedt expansion for $x_T(t)$ \cite{linstedt}. We seek to construct a 
solution of the oscillator equation (\ref{3}) with initial data $x(0)=
\epsilon>0$, $\dot{x}(0)=0$, by power series expansion in $\epsilon$. Such
a solution will be periodic in $t$ with unknown period $T(\epsilon)=
2\pi/\omega(\epsilon)$, where $T(\epsilon)$ and hence $\omega(\epsilon)$
depends analytically on $\epsilon$. 
Defining a rescaled time variable $s:=\omega(\epsilon)t$ (so the solution
is $s$-periodic with period $2\pi$) and dependent variable $y(s):=
x(t)/\epsilon$ (so $y(0)=1$, $y'(0)=0$ for all $\epsilon$), we substitute
the expansions
\bea
\label{30}
\omega(\epsilon)&=&a_0+a_1\epsilon+a_2\epsilon^2+\ldots \\
\label{31}
y(s)&=&y_0(s)+\epsilon y_1(s)+\epsilon^2y_2(s)+\ldots
\eea
into (\ref{3}) and Taylor expand $V(x)$ about $x=0$. Grouping terms by order
in $\epsilon$, this results in an infinite set of coupled ODEs for
$\{y_n:n\in\N\}$, to be solved in turn ($n=0,1,2,\dots$) with initial data
$y_0(0)=1$, $\dot{y}_0(0)=y_n(0)=\dot{y}_n(0)=0$. 
Demanding that each $y_n(s)$ have period $2\pi$ then
fixes the constants $a_0, a_1, a_2,\ldots$. 
For our purposes, it is sufficient
to proceed only up to order $\epsilon^2$, whereupon one finds that
\bea
\omega(\epsilon)&=&1-A\epsilon^2+O(\epsilon^3) \nonumber \\
\label{32}
\Rightarrow\qquad
T(\epsilon)&=&2\pi[1+A\epsilon^2+O(\epsilon^3)],
\eea
where $A=-V^{(4)}(0)/16-(V^{(3)}(0))^2/12$ is positive for each of our examples (since they
are all softly anharmonic, i.e.\ $T>2\pi$). Then 
\beq
\label{33}
x_{T(\epsilon)}(t)=\epsilon\cos\omega(\epsilon)t+\epsilon^2
y_1(\omega(\epsilon)t)+\epsilon^3 y_2(\omega(\epsilon)t)+\ldots
\eeq
where all we need know about $y_1,y_2,\ldots$ is that they
are even with period $2\pi$. We can now extract the asymptotic form
of $z_1(T(\epsilon))$ for small $\epsilon$:
\bea
z_1(T(\epsilon))&=&\int_0^{T(\epsilon)}\sin(T(\epsilon)-t)x_{T(\epsilon)}(t)
\, dt \nonumber \\
&=&\frac{1}{\omega(\epsilon)}\int_0^{2\pi}
\sin\left(\frac{2\pi -s}{\omega(\epsilon)}\right)(\epsilon\cos s+
\epsilon^2y_1(s)+\ldots)\, ds \nonumber \\
\label{34}
&=&A\pi^2\epsilon^3+O(\epsilon^4).
\eea
Hence from equation (\ref{14one}),
\beq
\label{35}
\chi_1(0)=\frac{z_1(T(\epsilon))}{1-\cos T(\epsilon)}
=\frac{1}{2\pi A\epsilon}+O(\epsilon^0).
\eeq
As $\epsilon\ra 0^+$, equations (\ref{32}) and (\ref{35}) reproduce the
$T\ra 2\pi^+$ asymptotic behaviour observed in all the numerical experiments.

\subsection{Vertical asymptotes}

We have seen that a sign change in $\chi_1(0)$, and hence a transition from
IPBs to APBs, must occur at $T=2n\pi$ ($n\in\Z^+$) unless $z_1(2n\pi)=0$, 
equation
(\ref{23}). In this section we will prove that, for a given $n$, the set of 
potentials on which (\ref{23}) holds is negligibly small, so that the
presence of asymptotes in the graphs of $\chi_1(0)$ against $T$ really is
generic. To be precise, we will show that, for potentials in a certain
Banach space $\mathcal{P}$, the subset of potentials on which (\ref{23})
holds is locally a codimension 1 submanifold.

\begin{defn} Let $\mathcal{P}=\{V:\R\ra\R\, \mbox{such that $V$ is $C^2$ and
$||V||_2<\infty$}\}$ and $F:\osc\oplus \mathcal{P}\ra\osco$ be the mapping
$$F(q,V)=\ddot{q}+V'(q).$$
Note that $F$ is a $C^1$ mapping between Banach spaces.
\end{defn}

\begin{defn} A potential $V\in \mathcal{P}$ is 
{\rm anharmonic at $q\in\osc$} 
 if $q$ is nonconstant,  $F(q,V)=0$, and $\ker{DF_{(q,V)}}=\{0\}$, where
$DF_{(q,V)}:\osc\ra\osco$ denotes the partial derivative of $F$ at $(q,V)$.
\end{defn}

Clearly $F(q,V)=0$ means that $q$ is a solution of Newton's equation 
for motion in potential $V$. That injectivity of $DF_{(q,V)}$ is equivalent
to the standard definition of anharmonicity in Hamiltonian mechanics is 
shown, for example, in the original proof of Theorem \ref{th:macaub}
\cite{macaub}. For all the potentials we considered, $V$ is anharmonic at
$q=x_T\in\osc$ for all $T>2\pi$.

\begin{defn} A {\rm perturbation
neighbourhood of $(q,V)\in\osc\oplus \mathcal{P}$} 
is a pair $(\mathcal{U},f)$ where $\mathcal{U} \subset \mathcal{P}$ is an 
open 
set
containing $V$, and $f:\mathcal{U}\ra\osc$ is a $C^1$ map satisfying 
(a) every $W\in \mathcal{U}$ is anharmonic at $f(W)$ and (b) $f(V)=q$. Note 
that
$\mathcal{U}$ is trivially a Banach manifold.
\end{defn}

\begin{lemma} For any $V\in \mathcal{P}$ anharmonic at $q\in\osc$ there 
exists a
perturbation neighbourhood $(\mathcal{U},f)$ of $(q,V)$. The function $f$ 
is unique on 
sufficiently small $\mathcal{U}$.
\end{lemma}
{\bf Proof:} Anharmonicity implies that $DF_{(q,V)}$ is injective, and 
hence the standard solvability criterion in linear ODE theory guarantees
it is also surjective (see section 3.3.2 of \cite{hasspe} 
for details). The open mapping 
theorem then ensures
boundedness of $DF_{(q,V)}^{-1}$.
Hence $DF_{(q,V)}$ is invertible, and the implicit function theorem applied
to $F$ ensures existence of $(\mathcal{U},f)$ and local uniqueness of $f$. 
$\Box$

\begin{defn} For $T\in 2\pi\Z^+$, 
we shall say that $V\in \mathcal{P}$ is {\rm
degenerate at $q\in\osc$} if $V$ is anharmonic at $q$ and
$$\int_0^T\cos\, t\, q(t)\, dt=0.$$
\end{defn}

\begin{thm} Let $T\in 2\pi\Z^+$, 
$V_0\in \mathcal{P}$ be degenerate at $q_0\in\osc$, and 
$(\mathcal{U},f)$ be a
perturbation neighbourhood of $(q_0,V_0)$. The subset of $\mathcal{U}$ on 
which 
degeneracy persists is a codimension 1 submanifold.
\end{thm}
{\bf Proof:} Consider the $C^1$ mapping $I:\mathcal{U}\ra\R$ defined by
$$I(V)=\int_0^T\cos\, t\, [f(V)](t)\, dt$$
($I$ is $C^1$ since $f$ is $C^1$). By the Regular Value Theorem 
(see appendix)
the result follows if we establish that $0$ is a regular value of $I$, or in 
other words that $I$ is a submersion at every $V\in I^{-1}(0)$.
Since ker $DI_V$ is of finite codimension in $T_V\mathcal{U}=
\mathcal{P}$ it splits and 
hence
it suffices to show that 
$DI_V:\mathcal{P}\ra\R$ is surjective for all $V\in I^{-1}(0)$.

Let $V\in I^{-1}(0)$, $q=f(V)$.  For any $\delta V\in \mathcal{P}$, let
$\delta q_{\delta V}=-[DF_{(q,V)}^{-1}(\delta V'\circ q)]\in\osc$. In other 
words 
$\delta q_{\delta V}$ is the unique even, $T$-periodic $C^2$ solution of
$$
\ddot{\delta q}_{\delta V}+V''(q(t))\delta q_{\delta V}=-\delta V'(q(t)),
$$
whose existence follows from anharmonicity of $V$ at $q$ (invertibility
of $DF_{(q,V)}$ on $\osco$). With this notation,
$$
DI_V(\delta V)=\int_0^T\cos t\, \delta q_{\delta V}(t)\, dt
$$
and surjectivity of $DI_V$ will follow if we exhibit $\delta V \in 
\mathcal{P}$
such that $DI_V(\delta V) \ne 0$. We construct such a $\delta V$ in two 
steps.

For some $t_0\in(0,T/2)$ and $\epsilon>0$, let $b\in\osc$ be a non-negative
function with $\mbox{supp}\, b\cap[0,T/2]=[t_0-\epsilon,t_0+\epsilon]$. 
Clearly
$$
\int_0^Tb(t)\cos t\, dt =  2\int_0^{T/2}b(t)\cos t\, dt\neq 0
$$
(where $T\in 2\pi\Z^+$ has been used) 
provided we choose $t_0\notin\cos^{-1}(0)$ and $\epsilon$ sufficiently
small. Since the critical points of $q(t)$ are isolated, we may also
assume that $q$ is invertible on $[t_0-\epsilon,t_0+\epsilon]$ with
$C^2$ inverse $q^{-1}:[x_1,x_2]\ra[t_0-\epsilon,t_0+\epsilon]$. 
By its construction the function $g$ defined by
$$
g(x)=\left\{\begin{array}{cc}
-[[DF_{(q,V)}b]\circ q^{-1}](x) & x\in[x_1,x_2] \\
0 & x\notin [x_1,x_2]
\end{array}\right. 
$$ 
is in $C^0(\R)$ (since $DF$ maps to $\osco$).
This allows us to construct a function $\delta \tilde{V}(x) \in C^1(\R)$ 
such that
$\delta q_{\delta \tilde{V}}=b$ as follows. For $x \in {\rm ran}(q)$, 
$\delta \tilde{V}(x)=\int_0^xg(z)\, dz$ defines a $C^1$ function on 
ran($q$). Since 
$\delta q_{\delta \tilde{V}}$ is clearly independent of the behaviour of 
$\delta \tilde{V}$ outside
${\rm ran}(q)$, we extend $\delta \tilde{V}$ to a $C^1$ function on $\R$
with support contained in some compact set $K \supset {\rm ran}(q)$.

We cannot immediately conclude that $DI_V$ is surjective, since $\delta 
\tilde{V}$
is $C^1$ but not necessarily $C^2$. However, using a density argument it is 
straightforward
to find a $\delta V \in C^2$ close enough to $\delta \tilde{V}$ such that 
$DI_V(\delta V) \neq 0$ stills holds.
More precisely, $DI_V$ has a natural continuous extension, 
call it $\ol{DI_V}$ to $C^1$. By construction $\delta \tilde{V}$ 
satisfies $\ol{DI_V}(\delta \tilde{V})\neq 0$ and belongs to $C_K^1$ 
(the $C^1$ functions with support contained in $K$).
Since $C^2_K$ is dense in $C^1_K$, 
there exists a sequence $\delta V_m\in C^2_K\subset \mathcal{P}$ 
converging (in $C^1$ norm) to $\delta \tilde{V}$, and by continuity  
$\ol{DI_V}(\delta V_m) \neq 0$ for all $m$ sufficiently large.
$\Box$


In the above, we have chosen $\mathcal{P}=(C^2(\R),||\cdot||_2)$ as our
space of potentials. This choice was made for the sake of clarity and 
notational 
simplicity -- many other choices would work. In fact, all but two ($V_{FK}$
and $V_G$) of the example potentials considered in section \ref{num} lie,
strictly speaking, outside $\mathcal{P}$, since they are unbounded. However,
the analysis above can easily be adapted to deal with this: one simply 
replaces $\mathcal{P}$ by the affine space $\mathcal{P}_V:=\{W:||W-V||_2<
\infty\}$. 

\section{Concluding remarks}
\label{con}

In this paper, breather initial profiles in the weak coupling regime of a
simple class of oscillator networks have been examined, focusing on the
dependence on breather period $T$. The direction of continuation of one-site
breathers was determined numerically using a simple and inexpensive
numerical scheme. Two types of breather were identified, called IPBs and
APBs. The numerical data suggest that generically these two types occur in
alternating bands in the $T$ parameter space ($T\in(2\pi,\infty)$), and that
the resonant periods $T\in 2\pi\Z^+$ separate an IPB band from an APB band.
The genericity of this behaviour was proved rigorously.

The distinction between IPBs and APBs, which
 does not appear to have attracted much attention
in the literature, may have phenomenological
implications in applications of the model (\ref{1}). One reason for interest
in discrete breathers (particularly continued one-site breathers)
 is that, because of their strong spatial localization, 
they typically require
little energy to achieve large amplitude oscillations close to the centre.
This makes them good candidates for ``seeds'' of mechanical breakdown of the
network, the idea being that the central oscillation becomes so violent as
to break the chain (a similar mechanism is postulated as a mechanism for
DNA denaturation, for example \cite{dna}). Since APBs stress the central
intersite springs more than IPBs they presumably make more effective seeds
of mechanical breakdown.

\section*{Acknowledgments}

This work was partially completed during a visit by MH to the 
Max-Planck-Institut f\"{u}r Mathematik in den Naturwissenschaften, Leipzig,
where JMS was a guest scientist. Both authors wish to thank Prof.\
Eberhard Zeidler
for the generous hospitality of the institute. JMS is an EPSRC Postdoctoral
Research Fellow in Mathematics. 
 
\section*{Appendix}

We recall some basic definitions and an elementary result of
infinite dimensional differential topology, the Regular Value Theorem, which 
gives conditions under which the level set of a smooth function is 
guaranteed to be a manifold.

\begin{defn}
We say that a closed subspace $S$ of a complete topological vector space 
$B$ {\rm splits} if there
exists another closed subspace $C$ which is complementary to $S$, i.e. 
$C + S = B$ and 
$C \cap S = (0)$.
\end{defn}
Note that if $B$ is a Hilbert space then any closed subspace $S$ splits since 
we can 
take the complement to be $C=S^{\perp}$. Also if $B$ is finite dimensional 
then since any subspace is closed all subspaces split. More generally any 
finite dimensional subspace 
(necessarily closed) of a Banach space splits, as does any closed subspace
of finite codimension.

\begin{defn}
  A $C^1$ map $f:X\ra Y$ between Banach manifolds is a {\rm submersion} at 
$x \in X$ if 
$Df_x:T_xX \ra T_{f(x)}Y$ is surjective and ker $Df_x$ splits. A value 
$y \in Y$ is a
{\rm regular value} of $f$ if $f$ is a submersion for every 
$x \in f^{-1}(y)$.
\end{defn}
Note that
in the case that $Y$ is finite dimensional then ker $Df_x$ is of finite 
codimension
in $T_x X$ and hence is guaranteed to split. In this case we need only 
verify that
$Df_x$ is surjective for $f$ to be a submersion at $x$.
\begin{thm}
  (The Regular Value Theorem) Let $f:X \ra Y$ be a $C^1$ map between Banach 
manifolds
and $y \in Y$ be a regular value of $f$. Then $f^{-1}(y)$ is a submanifold 
of $X$ 
with $T_x f^{-1}(y) \cong \mathrm{ker}\  Df_x$.
\end{thm}
The proof follows from working in charts around $x$ and $f(x)$ and applying 
the 
Implicit Function Theorem. For details of a proof see \cite{chobru2}.

\end{document}